\documentclass[a4paper]{jpconf}

\usepackage[english]{babel} 

\usepackage[%
	final,%
]{graphicx}

\usepackage[
   centertags, 
   sumlimits,  
   intlimits,  
   namelimits, 
]{amsmath} %
\usepackage{amssymb}

\usepackage[%
	pdftitle={Mach Cones in Viscous Matter},%
	pdfauthor={I. Bouras, A. El, O. Fochler, F. Lauciello, F. Reining, J. Uphoff,
		   C. Wesp, E. Moln\'ar, H. Niemi, Z. Xu and C. Greiner}
	pdfsubject={Investigation of shocks in viscous matter},
	pdfstartview=FitH,
	pdfpagemode=UseNone,
	bookmarksopen=true
	]{hyperref}

\usepackage[all]{hypcap} 

\usepackage{dcolumn} 

\usepackage{multirow} 

\usepackage{microtype} 

\makeatletter
\def\tagform@#1{\maketag@@@{\ignorespaces#1\unskip\@@italiccorr}}
\let\orgtheequation\theequation
\def\theequation{(\orgtheequation)}
\makeatother

\newcommand{\beq}{\begin{equation}}
\newcommand{\eeq}{\end{equation}}

\begin{document}
\title{Mach Cones in Viscous Matter}

\author{I. Bouras$^1$, A. El$^1$, O. Fochler$^1$, F. Lauciello$^1$, F. Reining$^1$,
J. Uphoff$^1$, C. Wesp$^1$, E. Moln\'ar$^{2,3}$, H. Niemi$^2$, Z. Xu$^{1,2}$ and C. Greiner$^1$}

\address{$^1$ Institut f\"ur Theoretische Physik, Johann Wolfgang Goethe-Universit\"at,
Max-von-Laue-Str.\ 1, D-60438 Frankfurt am Main, Germany}

\address{$^2$ Frankfurt Institute for Advanced Studies, Ruth-Moufang-Str. 1, D-60438 Frankfurt am Main, Germany}

\address{$^3$ KFKI, Research Institute of Particle and Nuclear Physics, H-1525 Budapest, P.O.Box 49, Hungary}

\ead{bouras@th.physik.uni-frankfurt.de}

\begin{abstract}
Employing a microscopic transport model we investigate the evolution of
high energetic jets moving through a viscous medium. For the scenario of
an unstoppable jet we observe a clearly strong collective behavior for a
low dissipative system  $\eta/s \approx 0.005$, leading to the observation of
cone-like structures. Increasing the dissipation of the system to
$\eta/s \approx 0.32$
the Mach Cone structure vanishes. Furthermore, we investigate jet-associated
particle correlations. A double-peak structure, as observed
in experimental data, is even for low-dissipative systems not supported,
because of the large influence of the head shock.
\end{abstract}

\section{Introduction}
The elliptic flow coefficient $v_2$ measured in heavy-ion collisions
at the Relativistic Heavy Ion Collider has a large value, leading
to the indication that the created quark-gluon plasma (QGP) behaves like
an almost perfect fluid \cite{Adler:2003kt,Adams:2003am}.
This is confirmed by recent calculations of viscous hydrodynamics
\cite{Luzum:2008cw} and microscopic transport calculations
\cite{Xu:2007jv,Xu:2008av}. Jet-quenching
\cite{Adams:2003kv} has been discovered and very exciting
jet-associated particle correlations have been observed
\cite{Wang:2004kfa}-\cite{Ulery:2005cc}.
They indicate the formation of shocks in form of Mach Cones induced by
highly-energetic partons traversing the QGP
\cite{Stoecker:2004qu}.
Measurements of the Mach Cone angle could give us the possibility to extract
the equation of state of the QGP.

In this work we address the question, under what conditions Mach Cones can
develop in viscous gluon matter for given constant $\eta/s$ values. Furthermore,
we discuss a possible double-peak structure in jet-associated
two-particle correlations.

\section{The parton cascade BAMPS}
\label{model}
The Boltzmann Approach of Multi Parton Scatterings (BAMPS) employed
in this study is a microscopic transport model solving the Boltzmann equation
\begin{equation}
p^{\mu} \partial_{\mu} f(x,p) = C(x,p)
\end{equation}
for on-shell particles with the collision integral $C(x,p)$.
The algorithm for collisions is based on the stochastic interpretation 
of the transition rate \cite{Xu:2004mz,Xu:2007aa}. In this study, we consider
only binary gluon scattering processes with an isotropic constant cross section,
which is related to the $\eta/s$ ratio via a simple
relation \cite{Xu:2007ns,El:2008yy}.

\section{Shocks Waves and Mach Cones}
Mach Cones, which are special phenomena of shock waves, have their origin
in ideal hydrodynamics \cite{Landau_book}. A very weak perturbation in a
perfect fluid induces sound waves which propagate with the speed
of sound $c_s = \sqrt{dp/de}$, where $p$ is the pressure and $e$ is the
energy density. In the case where the perturbation with velocity $v_{\rm jet}$
moves faster than the sound waves, the sound waves lie on a cone with an
emission angle $\alpha_w = \arccos ( c_s / v_{\rm jet} )$. Considering
a massless gluon gas, with $e = 3p$ and $c_s = 1/\sqrt{3}$, and a
perturbation with $v_{\rm jet} = 1$, the emission angle is
\begin{equation}
\label{eq:IdealmachAngle_wp_input}
\alpha_w = 54,73^\circ \, .
\end{equation}
In the case of stronger perturbations, the sound waves move
faster than the speed of sound through the medium and therefore are
called shock waves \cite{Rischke:1990jy}. Then we can approximate
the emission angle by
\begin{equation}
\label{eq:IdealMachAngle}
\alpha \approx \arccos \frac{ v_{\rm shock} }{v_{\rm jet}} \, .
\end{equation}
The velocity of the shock front $v_{\rm shock}$ depends on
the pressure (energy density) in the shock front region
$p_{ \rm 0}$ ($e_{ \rm 0}$) and in the stationary medium itself
$p_{ \rm 1}$ ($e_{ \rm 1}$):
\begin{equation}
\label{eq:v_shock}
v_{\rm{shock}} = \left [ \frac{(p_1 - p_0)(e_0 + p_1)}
{(e_1 - e_0)(e_1 + p_0)} \right]^{\frac{1}{2}}
\end{equation}
Eq.\eqref{eq:v_shock} has the following limits: For
$p_0 >> p_1$ we obtain $v_{\rm{shock}} = 1$. If
$p_0 \approx p_1$, that is, a very weak perturbation,
we get the expected limit of the speed of sound
$v_{\rm{shock}} \approx c_s$. In the latter case
Eq.\eqref{eq:IdealMachAngle} reduces to the one
given by $\alpha_w$, as expected.

\section{Transition from ideal to viscous Mach Cones in BAMPS}
In this section we employ the microscopic transport model BAMPS to
investigate Mach Cones with different strength of dissipations in the medium.
All simulations are realized within a static and uniform medium of
massless Boltzmann particles and $T = 400$ MeV. To save computational
runtime we reduce our problem to two dimensions. The physical results
compared to full three-dimensional simulations are similar except for
the fact that the energy density decreases slower in our reduced
geometry. Here we choose the $xz$-plane and apply a periodic boundary
condition in $y$-direction.

A jet moving in positive $z$-direction is initialised at $t = 0$ fm/c at
the position $z = -0.8$ fm. The jet is treated as a massless particle with zero
spatial volume and zero transverse momentum, that is, $p_z = E_{\rm jet} = 200$
GeV and $v_{\rm jet} = 1$. The jet deposits energy and momentum to the
medium via collisions with medium particles. In this scenario we neglect
the deflection of the jet; its energy and momentum is set to its initial
value after every collision.

%
\begin{figure}[th]
\includegraphics[width=\columnwidth]{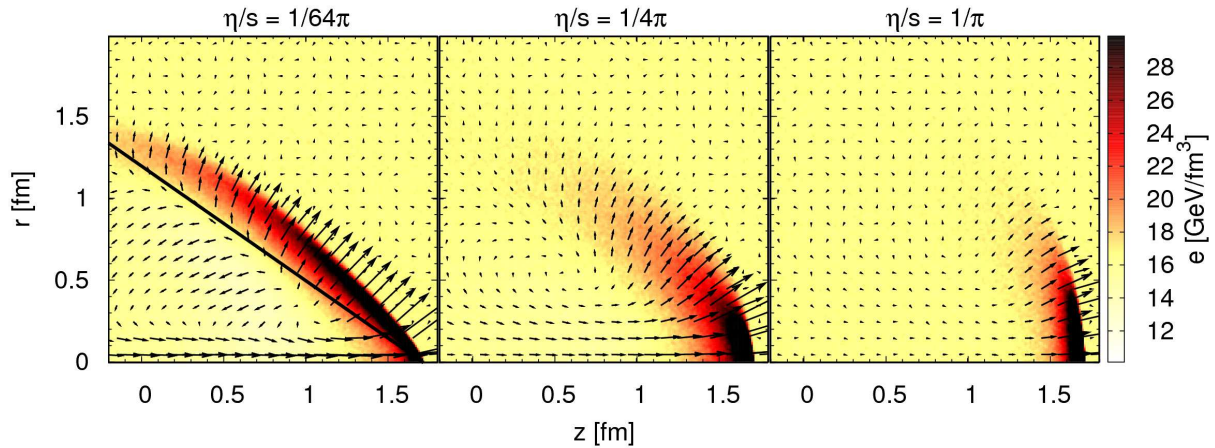}
\caption{(Color online) The shape of a Mach Cone shown
for different viscosities of the medium,
$\eta/s = 1/64\pi$ (left), $\eta/s = 1/4\pi$
(middle), $\eta/s = 1/\pi$ (right). We show here
the energy density plotted together with the velocity profile.
Additionally, in the left panel the linear ideal Mach Cone for
a very weak perturbation is shown by a solid line; its emission
angle is \eqref{eq:IdealmachAngle_wp_input}.}
\label{fig:machCone}
\end{figure}
%

In Fig.\ref{fig:machCone} we demonstrate the transition from ideal
Mach Cone to a highly viscous one by adjusting the shear viscosity
over entropy density ratio in the medium from
$\eta/s = 1/64 \pi \approx 0.005$ to $1/\pi \approx 0.32$.
The energy deposition of the jet is approximately $dE/dx = 11 - 14$
GeV/fm. The simulations are shown at $t = 2.5$ fm/c.

Using a non-physical small viscosity of $\eta/s = 1/64\pi$ we
observe a strong collective behavior in form of a Mach Cone,
as shown in the left panel of Fig.\ref{fig:machCone}.
For comparison, the ideal Mach Cone caused by a very weak
perturbation is given by a solid line. Its emission angle
is given by \eqref{eq:IdealmachAngle_wp_input}. The shock wave,
characterized by a higher energy density compared to the
medium at rest, propagates with the emission angle $\alpha$ smaller
than the ideal one, $\alpha_w$. Due to the fact that the energy deposition
is strong, the shock propagates through the medium faster than the
speed of sound, given approximately by Eq.~\eqref{eq:IdealMachAngle}.
A simulation where the energy deposition to the medium is much
smaller reproduces the ideal Mach Cone with its emission angle
$\alpha_w$ \eqref{eq:IdealmachAngle_wp_input}.

Furthermore, a strong diffusion wake in direction of the jet,
characterized by decreased energy density, and a head shock in
the front are clearly visible. The results agree qualitatively
with those found in \cite{Bouras:2010nt}-\cite{Molnar:2009kx}.

If we increase the viscosity of the medium to larger values, shown
in the middle and left panel of Fig.\ref{fig:machCone}, the
typical Mach Cone structure smears out and vanishes completely.
Due to stronger dissipation, the collective behavior gets weaker
because of less particle interactions in the medium with a larger
$\eta/s$. The results agree qualitatively with the analysis of
one-dimensional shock waves realized in earlier studies using
kinetic theory and viscous hydrodynamics
\cite{Bouras:2009nn}-\cite{Bouras:2010hm},
where a smearing-out of the shock profile is observed with
higher viscosity.

\section{Two-particle Correlation}
\begin{figure}[th]
\begin{minipage}[l]{0.5\textwidth}
\includegraphics[width=\textwidth]{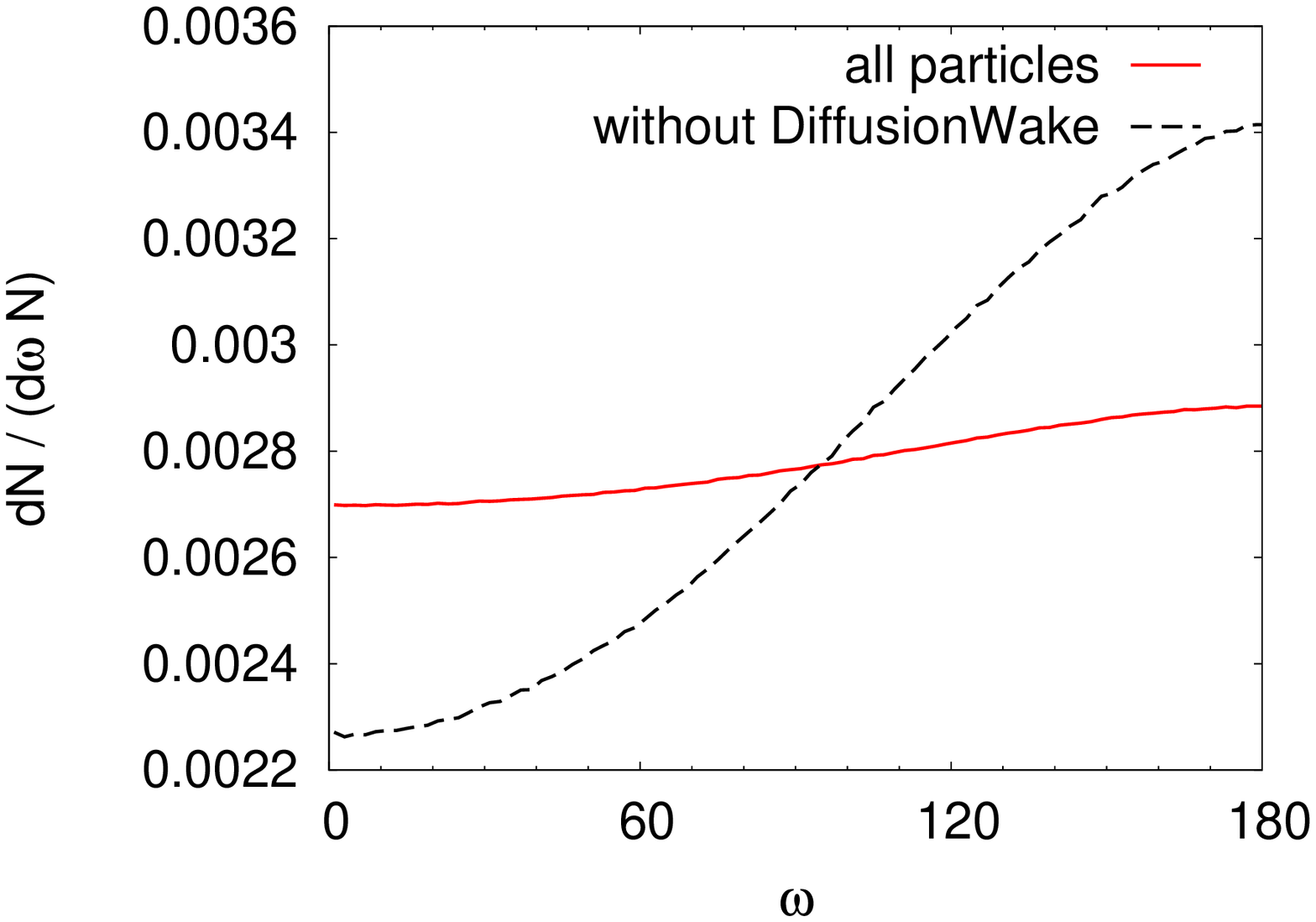}
\end{minipage}
\begin{minipage}[r]{0.5\textwidth}
\includegraphics[width=\textwidth]{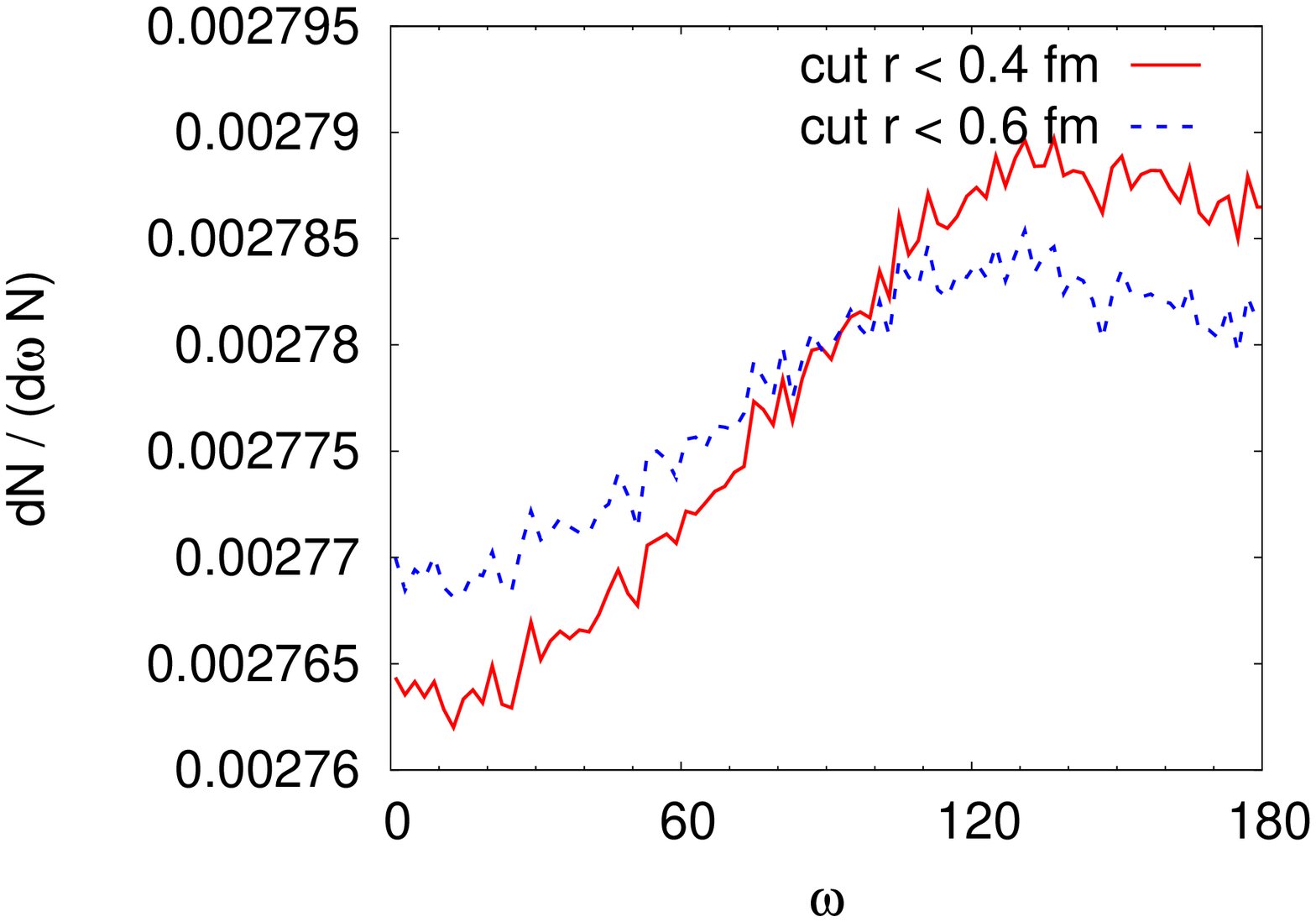}
\end{minipage}
\caption{(Color online) Jet-associated particle correlations
extracted from the simulation with $\eta/s = 1/64\pi$ as discussed
in Fig.\ref{fig:machCone}. The left panel shows
the correlations including all particles and one without the
part of the diffusion wake. The right panel shows the correlations
by systematically cuts in space to subtract the contribution of
the diffusion wake and head shock. $r$ is the transverse
direction to the $z$-axis.}
\label{fig:two_particle}
\end{figure}

The existence of a conical structure in Fig.\ref{fig:machCone} for
low viscosity are very exciting and naively lead to the assumption
of a double-peak structure in jet-associated particle correlations.
In Fig.\ref{fig:two_particle} we show the particle distribution
$dN/(d\omega N)$ plotted vs. $\omega$ as extracted from BAMPS, where
$\cos \omega = p_z/\sqrt{p_x^2 + p_z^2}$. We remark at this point that
one has to mirror the results for $180^\circ$ - $360^\circ$.

%
%

In the left panel we show the particle distribution for all
particles. A peak in direction of the jet at $180^\circ$ is observed.
A big contribution of the diffusion wake and head shock is
present and hinders therefore the appearance of a double peak structure.
In the same figure the particle distribution without the diffusion wake
is plotted. However, the contribution of the head shock is still
too large to get a double peak structure. A systematic cut
in space is shown in the right panel, where we try to cut simultaneously
a part of the diffusion wake and head shock. As one can see, a
double peak structure appears and is enhanced with larger cut in
space. However, such a cut in space does not relate to any experimental
observation in heavy ion collisions and has to be treated with caution.

We conclude that a double peak structure exists, but the
contribution of the diffusion wake and head shock might be stronger.
This phenomena will be discussed in future studies.

\section*{Acknowledgements}

The authors are grateful to the Center for Scientific 
Computing (CSC) at Frankfurt University for the computing resources.
I.\ B. is grateful to HGS-Hire.
E.\ M. acknowledges support by OTKA/NKTH 81655 and
the Alexander von Humboldt foundation. The work of H.\ N. was supported by
the Extreme Matter Institute (EMMI).

This work was supported by the Helmholtz International Center
for FAIR within the framework of the LOEWE program 
launched by the State of Hesse.

\section*{References}


\end{document}